\documentclass[]{aa}
\usepackage{graphicx,epsf,psfig,epsfig,times,amsmath,amssymb}

\usepackage{natbib}
\bibpunct{(}{)}{;}{a}{}{,}

\begin{document}
\title{Resolving the hard X-ray emission of GX 5-1 with \textit{INTEGRAL}\thanks{Based on
observations with \textit{INTEGRAL}, an ESA project
with instruments and science data centre funded by ESA member states
(especially the PI countries: Denmark, France, Germany, Italy,
Spain, and Switzerland), Czech Republic and Poland, and with the
participation of Russia and the USA.}}
\author{A. Paizis\inst{1,}\inst{2},  K. Ebisawa \inst{3,}\inst{1}, T. Tikkanen \inst{4},
J. Rodriguez \inst{5,1}, J. Chenevez\inst{6}, E. Kuulkers\inst{7},   
O. Vilhu\inst{4} \\
and  T.J.-L. Courvoisier\inst{1,}\inst{8}   }
\offprints{Ada.Paizis@obs.unige.ch}

\institute{\textit{INTEGRAL} Science Data Centre, Chemin d`Ecogia 16, 1290 Versoix, Switzerland
\and
CNR-IASF, Sezione di Milano, Via Bassini 15, I--20133 Milano, Italy
\and
NASA Goddard Space Flight Center, Code 662, Greenbelt, MD 20771, USA
\and
Observatory, P.O.Box 14, T\"ahtitorninm\"aki, Fi--00014 University of Helsinki, Finland
\and
CNRS, FRE 2591, CE Saclay DSM/DAPNIA/SAp, F--91191 Gif sur Yvette Cedex, France
\and
Danish Space Research Institute, Juliane Maries Vej 30, DK--2100 Copenhagen, Denmark
\and
Research and Scientific Support Department of ESA, ESTEC, P.O. Box
299, NL--2200 AG Noordwijk, The Netherlands
\and 
Observatoire de Gen\`eve, 51 chemin des Mailletes, CH--1290 Sauverny, Switzerland
}
\date{Received  / Accepted}
\authorrunning{Paizis et al.}
\titlerunning{Resolving the hard X-ray emission of GX 5-1 with \textit{INTEGRAL}}

\abstract{We present the study of one year of  \textit{INTEGRAL} data on the neutron star low mass X-ray 
binary GX 5--1. Thanks to the excellent angular resolution and sensitivity of \textit{INTEGRAL}, 
we are able  to obtain a high
quality spectrum of GX 5--1 from $\sim$5\,keV to $\sim$100\,keV, for the first time  without contamination
from  the nearby black hole candidate \mbox{GRS 1758-258} above 20\,keV.
During our observations, GX 5--1 is mostly found in the horizontal and normal branch of
its hardness intensity diagram. A clear hard X-ray emission is observed 
above $\sim$30\,keV which exceeds the exponential cut-off spectrum expected from 
lower energies. This spectral flattening may have the same origin of the hard components observed in other Z sources
as it shares the property of being characteristic to the horizontal branch. 
 The hard excess  is explained by introducing  Compton up-scattering
of soft photons from the neutron star surface due to a thin hot plasma expected in the
boundary layer.
The spectral changes of GX 5--1 downward along the "Z" pattern  in the 
hardness intensity diagram can be well described in terms of  monotonical
decrease of the neutron star surface temperature. This may be a consequence of the
gradual expansion of the boundary layer  as the mass accretion rate increases.

}
\maketitle
%
%________________________________________________________________
\section{Introduction}
Low-mass X-ray binaries (LMXRBs) are systems where the compact object, either 
a neutron star (NS) or a black hole candidate (BHC), accretes matter 
from a companion with a mass  \mbox{$\textit{M}\lesssim$ $1\textit{M}_{\odot}$}.
LMXRBs hosting a weakly magnetised neutron star can be
broadly classified in two classes \citep{hasinger89}:
high luminosity/Z sources and low luminosity/Atoll sources.
In the colour-colour (CC) and X-ray hardness intensity  diagrams (HID), Atoll
sources display an upwardly curved branch while
Z sources describe an approximate ``Z'' shape.
Although two recent studies \citep{muno02,gierlinski02}
have suggested that the clear Z/Atoll distinction on the CC 
diagram is an artifact due to incomplete sampling (Atoll sources, if
observed long enough, \emph{do} exhibit a Z shape as well) many differences remain. 
Atoll sources have weaker
magnetic fields (about $10^6$ to $10^7$\,G versus 10$^8$--10$^9$\,G of
Z sources), are generally fainter ($0.01$--$0.3\,L_{\text{Edd}}$
versus $\sim$$L_{\text{Edd}}$), can exhibit harder spectra and trace out the Z shape on longer time
scales than typical Z-sources. Besides, their timing variability properties in a given spectral state are different.

The study of these sources in the hard X-ray domain has proven important especially 
in the recent years, in the light of the discovery of hard tails extending up to $\sim$100\,keV
in NS LMXRBs.
This kind of hard emission was thought to be a prerogative of BHCs and had been proposed as a possible signature
for the presence of a BH \citep[see e.g.,][]{mcclintock04}. Hard tails discovered  in about 20 NS LMXRBs of the 
Atoll class as well as in some Z sources showed that NS binaries as well can power such an emission \citep{barret94,disalvo02}.

GX 5--1 (4U 1758-25) is thought to be a NS LMXRB and is one of the six currently known Galactic 
Z sources\footnote{The other known
Galactic Z sources are GX340+0, GX349+2, Cyg X--2, GX 17+2 and Sco X--1.}.
It displays a complete Z pattern in the CC and HIDs 
\citep{kuulkers94, jonker02} and  shows secular shifts of the "Z". 
 As most of the Z sources, it is located near the Galactic Centre which has rendered 
difficult its study due to source confusion and optical obscuration.
Throughout the paper we assumed a distance of 8\,kpc and column density 
N$_{{\rm H}}$ = 3$\times$10$^{22}$ cm$^{-2}$  \citep{asai94}. 
Like all Z sources, GX 5--1 is a radio source \citep[e.g.][]{fender00} with the radio 
emission most likely originating in a 
compact jet.
The determination of the radio counterpart allows for extremely accurate position 
measurements. These have led to a most likely candidate for an infra-red 
companion \citep{jonker00}. 
In the X-ray domain, up to now neither pulsations nor bursts have been 
detected \citep[e.g.][]{vaughan94}. 

Previous GX 5-1 hard X-ray data were almost always contaminated by
the nearby (40$^{\prime}$) BHC LMXRB \mbox{GRS 1758-258}.  
The first mission to clearly resolve GX5-1 from \mbox{GRS 1758-258} in the high energy domain
was  \textit{GRANAT} \citep{paul91}.
Observations from  ART-P ($\sim$5--20\,keV), on board  \textit{GRANAT},
showed that  GX 5--1 was $\sim$30-50 times 
brighter than GRS 1758-258 below 20\,keV. At higher energies, SIGMA detected only 
GRS 1758-258 \citep{sunyaev91, gilfanov93}.

\begin{figure*}
\centering
\includegraphics[width=1.0\linewidth]{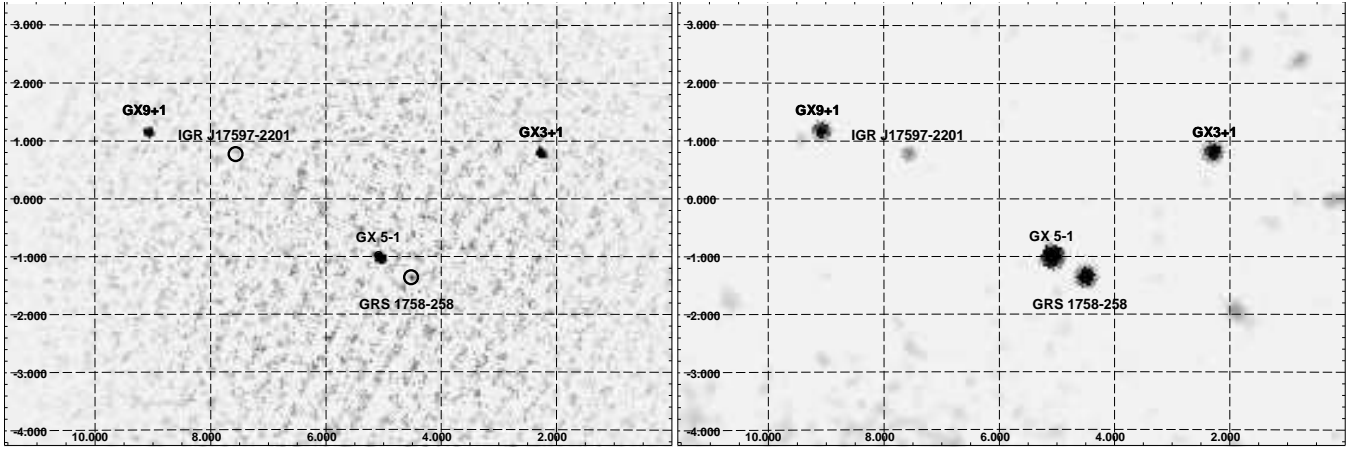}
%\vspace{4cm}
\caption{Mosaic images combining the April 2003 and October 2003 data sets. \emph{Left panel}: 
JEM--X mosaic image of GX 5--1 in the 5--10\,keV band (44 ScWs, $\sim$76\,ksec).
 \emph{Right panel}:
IBIS/ISGRI mosaic image of GX5--1 in the 20--30\,keV band (90 ScWs, $\sim$167\,ksec). The source IGR~J17597$-$2201 
 has been recently discovered by \textit{INTEGRAL} \citep{lutovinov03}.
 \label{fig:ima}}
\end{figure*}
\begin{figure*}
\centerline{
\begin{tabular}{cc}
\psfig{file=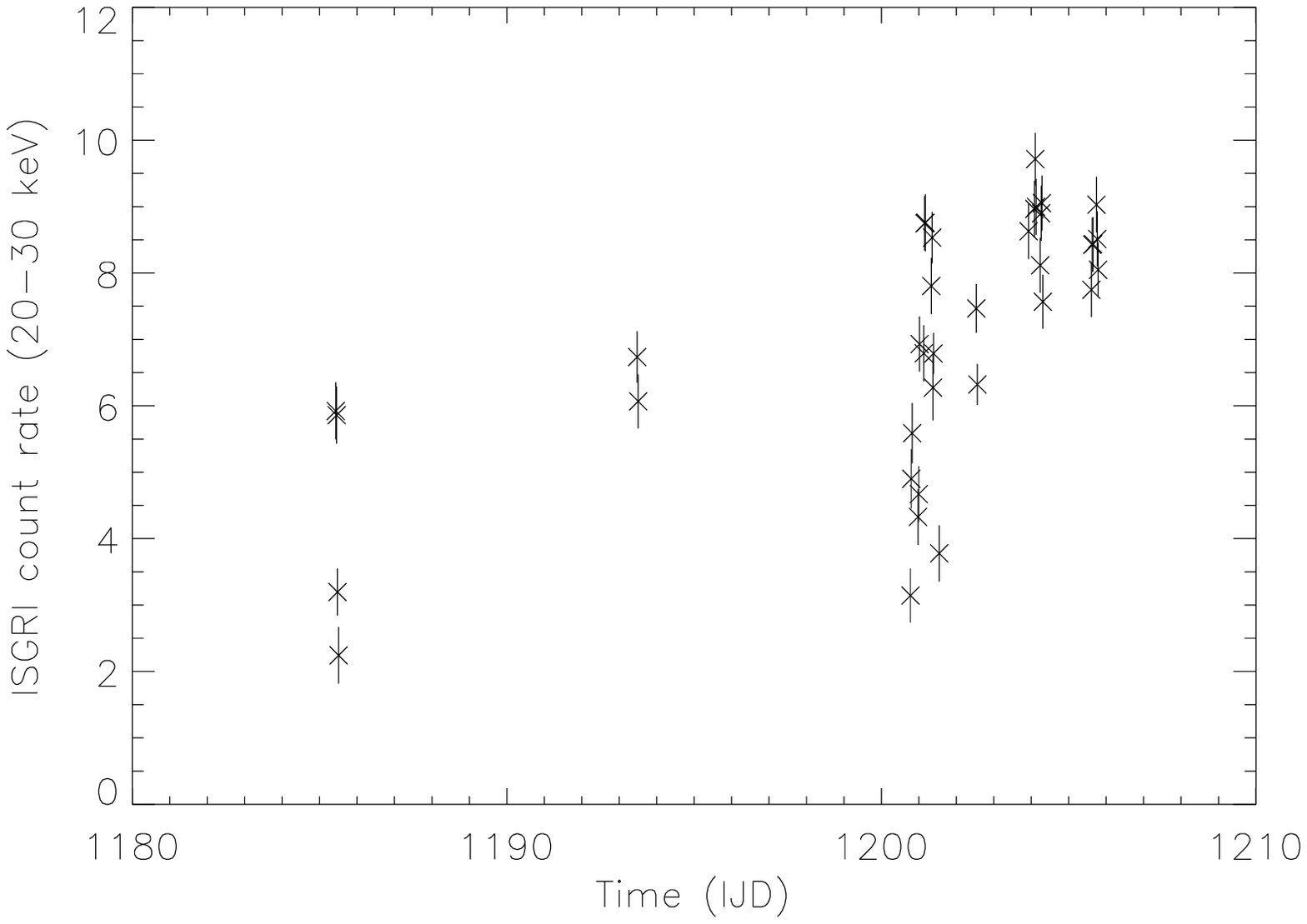, width=0.5\linewidth} &
\psfig{file=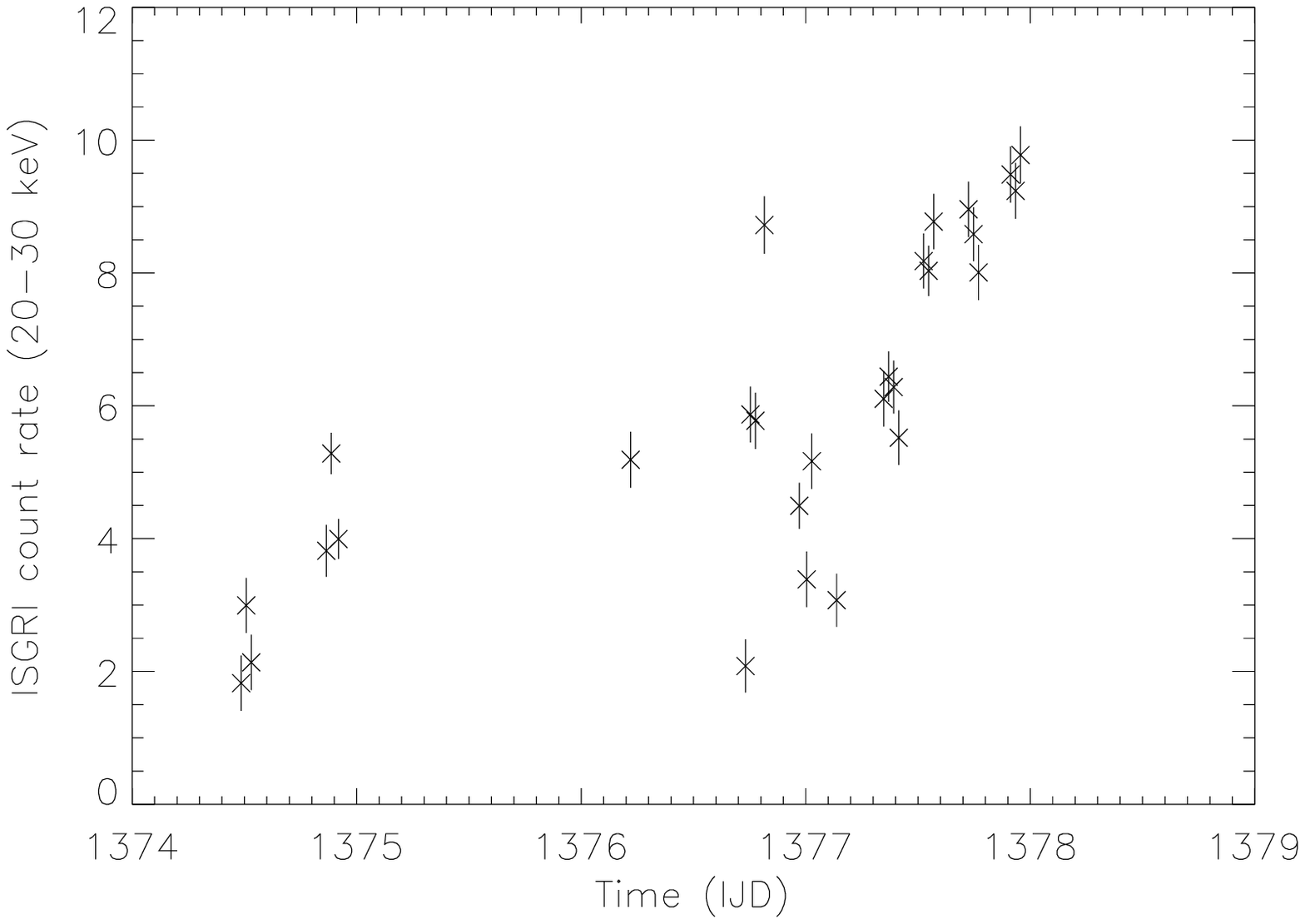, width=0.5\linewidth} 
\end{tabular}}
\caption{IBIS/ISGRI light curves of GX 5--1 in the 20--30\,keV for the April 2003 and October 2003 
coverage (left and rigth panel, respectively). Each point is one ScW (about 2000 s). Error bars are
at 1$\sigma$.
The observed countrate corresponds to a variability between about 30 and 130\,mCrab. In the softer band (i.e. in JEM--X) the source is much 
brighter, around 1 Crab.\label{fig:isgrilc}}
\end{figure*}

In this paper we report the results of one year of monitoring of GX 5--1 with the  
INTErnational Gamma-Ray Astrophysics Laboratory, \textit{INTEGRAL} \citep{winkler03}.
\textit{INTEGRAL} is an ideal mission to study the hard X-ray emission of sources in the 
crowded Galactic Centre region. \textit{INTEGRAL}/IBIS \citep{ubertini03} has high sensitivity, 
about $\sim$10 times better than SIGMA, coupled to imaging capability with 12$^{\prime}$ angular resolution above 20\,keV.
In addition, two X-ray monitors, JEM--X1 and JEM--X2 \citep{lund03}, provide the soft-X ray  
simultaneous coverage (3--35\,keV, 3$^{\prime}$ angular resolution). 
Effectiveness of \textit{INTEGRAL} in the study of Galactic bulge
bright LMXRBs has been demonstrated in  \cite{paizis03, paizis04} using only initial data.

\section{Observations and data analysis}

\textit{INTEGRAL} performed a Galactic Plane Scan (GPS) about every 12 
days and Galactic Centre Deep Exposures
(GCDEs)  according to the Galactic Centre visibility. 
Among these observation programs,  GX 5--1 has been continuously
in the \textit{INTEGRAL} field of view  around two distinct periods, April 2003 
(\textit{INTEGRAL} Julian Date\footnote{The \textit{INTEGRAL} Julian Date is defined as the 
fractional number of days since January 1, 2000 (TT). IJD=MJD-51544. } IJD$\simeq1200$,  
 hereafter data set 1, $\sim$90\,ksec)
and October 2003 (IJD$\simeq1380$, hereafter data set 2, $\sim$77\,ksec).

For the analysis we  used the available X-ray monitor, JEM--X2 (hereafter JEM--X), for 
 soft photons (3--35\,keV) and the low energy IBIS detector, ISGRI \citep{lebrun03}
for harder photons (20--200\,keV).  We have not
used PICsIT, the hard photon IBIS detector \citep{dicocco03}, as its 
peak sensitivity is above 200\,keV, well beyond the 20\,keV detection limit given by SIGMA.
The spectrometer SPI \citep{vedrenne03} has not been used as the 
presence of the nearby BH GRS 1758-258 could contaminate the data. 
The separation of the two sources, 40$^{\prime}$,
is below the angular resolution of the instrument, $\sim2.5^\circ$.  

JEM--X and IBIS/ISGRI data were reduced using the Off-line Scientific Analysis (OSA) 
package, OSA 4.0. The description of the algorithms used in the scientific analysis can be found in  
\cite{westergaard03} and  \cite{goldwurm03} for JEM--X and IBIS/ISGRI respectively.
XSPEC 11.3.0 was used for the spectral analysis. 
The Comptonisation model by \cite{nishimura86}  (COMPBB) 
was updated so that the cutoff energy of the model, originally at 70\,keV, would suit ISGRI spectra 
(i.e. moved up to 300\,keV).

We used JEM--X to build the hardness intensity diagram (HID) of GX 5--1. This was done using all the JEM--X data available with 
off-axis angle $<5^\circ$ which led to 113 science windows (hereafter ScWs) around 2000 sec exposure each\footnote{A science window is
the basic unit of an \textit{INTEGRAL} observation. Under normal operations one ScW will correspond to one
pointing or one slew. To build the HID, we considered 113 pointings, 13 from the GPS ($\sim$2200\,sec each) 
and 100 from the GCDE ($\sim$1800\,sec each).}.
 We extracted individual images from each  
ScW in 3 energy bands (3--5, 5--12 and 12--20\,keV).
A  mosaicking tool by \cite{chenevez04}
was used to obtain the final JEM--X  mosaic image.
Within each ScW we extracted 100 s bin light-curves 
and used one ScW and 100 s data bins to build the HID. 
For the spectral fitting of JEM--X 
data it is preferable to have the source within $3^\circ$ off-axis (better vignetting correction and 
signal-to-noise ratio). This led to 44 JEM--X ScWs. 
Spectra were grouped so that each new energy bin contained at least 200 counts. Based on Crab calibration results,
 3\% systematics have been applied  between 5 and 22\,keV were the spectral analysis has been performed.\\
In the case of ISGRI we selected the data within the totally coded field of view (FOV, $<$$4.5^\circ$, to 
avoid the partially to totally coded FOV edge),
resulting in a total of 90 ScWs (44 of which are in common with JEM--X).
We extracted individual ScW images in 10 energy ranges with the boundaries 
20, 30, 40, 50, 60, 80, 100, 150, 300, 500 and 1000\,keV. 
The fluxes extracted in each image and energy band were used to build the light-curves 
of GX 5--1. Images from each ScW were combined in a final mosaic
from which an overall ISGRI spectrum was extracted.
Spectra were also extracted from each ScW with the Least Square Method and 
 were further grouped so that each new energy bin had a  minimum of 20 counts. A 5\% systematics has been
   applied to all channels. For the spectral fitting we used the OSA 4.0
   ancillary response file (ARF) and redistribution matrix (RMF) re-binned to 
   16 spectral channels between 20 and 120\,keV.\\ 
Simultaneous ISGRI and JEM--X fitting was carried out for the 44 JEM--X ScWs. The remaining 46 ISGRI ScWs
were not analysed separately because  there was not enough statistics to 
study ISGRI spectra alone on a ScW basis.
In the JEM--X - ISGRI simultaneous fit a cross-calibration factor was let free to vary.

\section{Results}
\subsection{Overview of the data}
Figure~\ref{fig:ima} shows the JEM--X, 5--10\,keV, and IBIS/ISGRI, 20--30\,keV, mosaic image of the region 
of GX 5--1 (April 2003 and October 2003 data sets combined).
The nearby BHC \mbox{GRS 1758-258} is rather dim 
in the softer energy range (JEM--X) and becomes much brighter in the harder range (ISGRI) where it is 
clearly disentangled from GX 5--1. In the ISGRI mosaic, we detect GX 5--1 with about 4\,counts/sec and 
a detection significance of about 100. In the same image, \mbox{GRS 1758-258} is detected with an average countrate of about 2\,counts/sec and 
a detection significance of about 50.  For comparison,
the average count rate obtained from Crab, in different positions of the totally coded field of view, is about 75\,counts/sec in the 
20--30\,keV. 
In Fig.~\ref{fig:isgrilc} the overall ISGRI 20--30\,keV light curve of GX 5--1 is shown. Each data point
corresponds to one ScW in which GX 5--1 has been automatically detected by the software.
This happens in 66 ScWs (out of 90) in the 20--30\,keV band (with detected countrate between 2 and 10\,counts/sec, i.e. between about 
30 and 130\,mCrab) and in 
10 ScWs in the 30--40\,keV band (less than 2\,counts/sec, $\sim$60\,mCrab). In the remaining ScWs, \mbox{GX 5--1} was
not reaching the detection significance of 3, set as a threshold for the automatic detection in a single ScW. 
The combination of all the ScWs ($\sim$2\,ksec each) in a single 
mosaic (167\,ksec, Fig.~\ref{fig:ima}) allows to reveal a hard X-ray emission 
above $\sim$20\,keV. \\
In the softer energy range covered by JEM--X, \mbox{GX 5--1} is much brighter (around 1\,Crab) and is detected in each 100\,sec bin
with an average of 27\,counts/sec (3--5\,keV),  56\,counts/sec (5--12\,keV) and 
5\,counts/sec (12--20\,keV).

\subsection{The hardness intensity diagrams}
In order to build the HID of GX 5--1  we used the JEM--X soft colour, defined as 
the (5--12\,keV) to (3--5\,keV) flux ratio, plotted versus the JEM--X intensity (defined as 
the 3--12\,keV flux).

\begin{figure}
\centering
\includegraphics[width=1.0\linewidth]{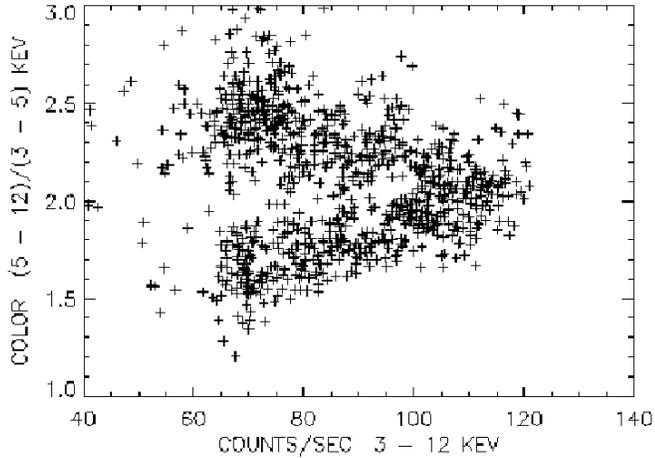}
%\vspace{4cm}
\caption{HID of GX5--1 with JEM--X (data set 1). The soft colour is the (5--12\,keV) to (3--5\,keV) flux ratio.
The graph was built using 100 s time bins that were smoothed with a three bin box-car: this means that
each 100 s data value is replaced by the mean of the three neighbouring values. The smoothing was chosen
for visualisation purposes as it makes the "Z" more evident. The horizontal branch (upper part, soft colour
$>2$)  and normal branch (lower part,  soft colour $<2$) are clearly visible.
\label{fig:Zsmooth}}
\end{figure}
Figure~\ref{fig:Zsmooth} shows the HID we obtained for the first data set (April 2003). The graph was built starting from
the 100 s bins that were smoothed to a 5 minute final bin for visualisation purposes. 
The horizontal branch, HB, and normal branch, NB, are quite evident while the flaring branch, FB, the last part of the "Z",
is not visible. The FB could be hidden in the data since in GX 5--1 it is short and/or spread out by our choice of energy boundaries. 
Note the similarity with the HID derived from \textit{GINGA}/ASM data
\citep{vanderklis91} and \textit{Mir/Kvant} TTM data \citep{blom93}.

\begin{figure*}
\centerline{
\begin{tabular}{cc}
\psfig{file=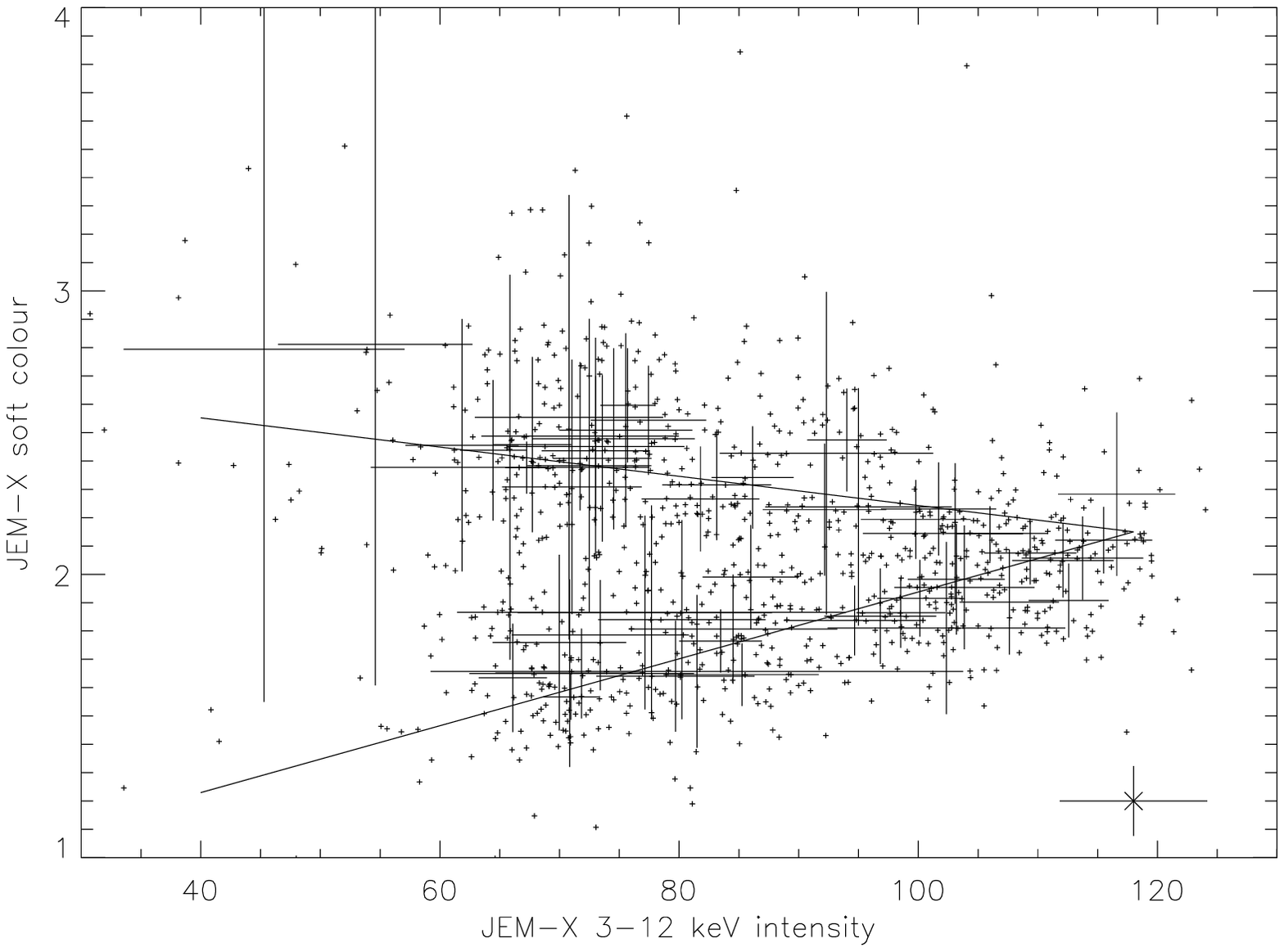, width=0.5\linewidth} &
\psfig{file=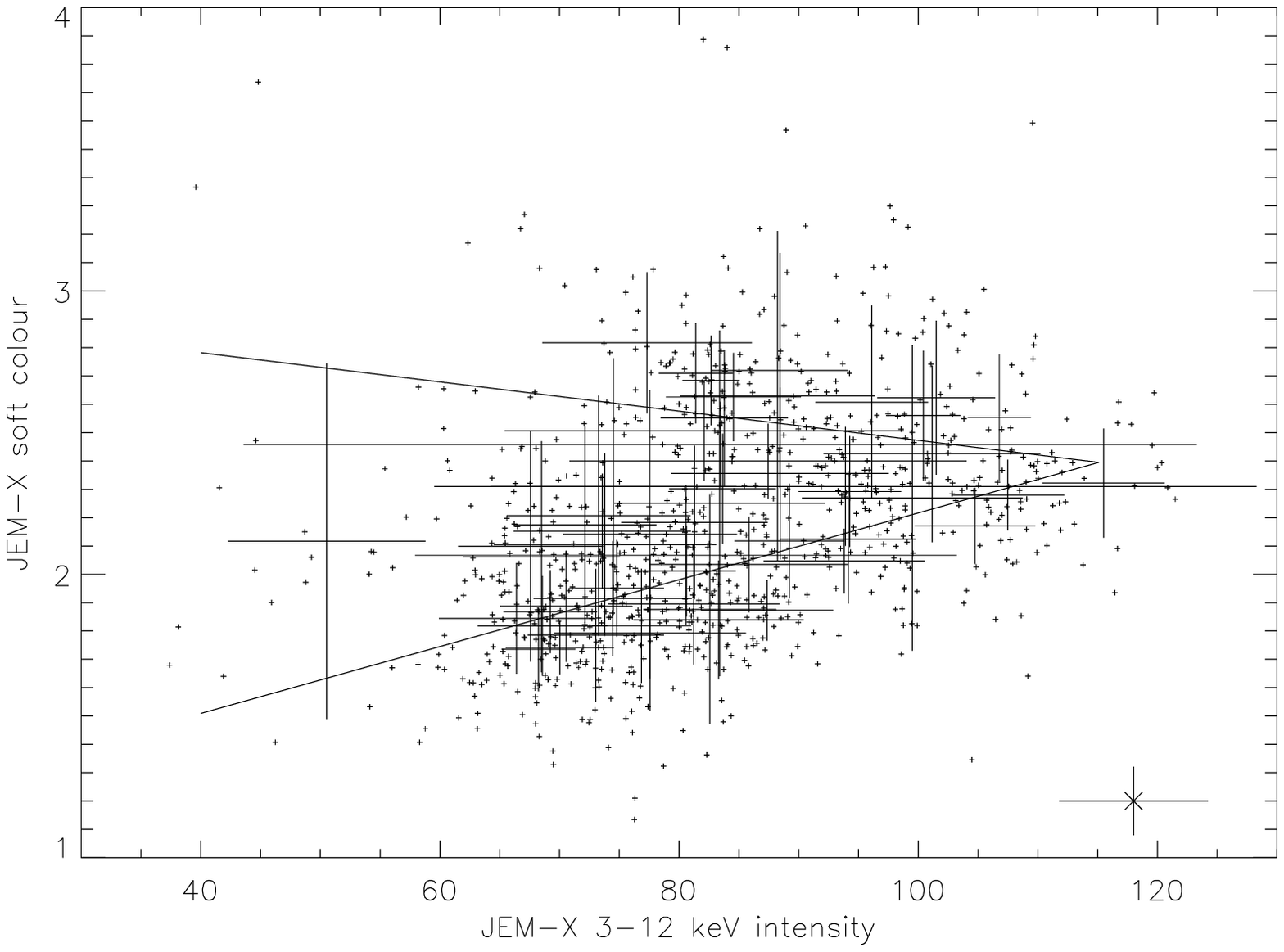, width=0.5\linewidth} 
\end{tabular}}
\caption{HID of GX5--1 with JEM--X for the two data sets.  
The scattered data points are 100 s bins while the points with error bars are one ScW bins. 
The average statistical error of the 100 s bins is shown in the 
bottom right of the graphs. The ScW bin values are computed as average of 100 s bins and the associated error
is not the statistical one but a measure of how much the source moved along the "Z" track within the ScW.  The solid line 
shows the deduced "Z" for the two data sets. 
\emph{Left panel}: data set 1 (April 2003). \emph{Right panel}: data set 2 (October 2003). The "Z" shape is less
evident in this data set with respect to the left panel. This change could be intrinsic to the source as the 
exposure time in the two data sets is comparable (58 ScWs in data set 1 and 55 ScWs in data set 2).  \label{fig:Z}}
\end{figure*}

Figure~\ref{fig:Z} shows the "raw", i.e. non-smoothed, HIDs obtained for the two data sets 
(April 2003 and October 2003 respectively).
 In the first data set the HB-NB vertex 
was set to (118,2.15) judging from the overall distribution of the data points. Then the slopes of the "Z" were 
calculated so that an equal number of data points were left on either side of the line. The deduced slopes are shown in 
Fig.~\ref{fig:Z} with a solid line.
As the first data set has a cleaner "Z" pattern, we used this one as the "true" one to 
which the second "Z" was forced to bend. This was done shifting the HB-NB vertex 
of the second data set until the slopes of HB and NB became 
equal to those of the first data set. This led to a vertex in the second set at (115.1, 2.4) which
resulted in a clear secular shift from 2.15 to 2.4 in the soft colour of the vertex.
Since the current data have an almost continuous coverage from the HB through the 
NB while coverage of the FB (if any) is very poor, in this paper we focus on the HB and NB.

In order to investigate the correlation between the source position in the "Z" pattern of the HID and the spectral
behaviour, we introduced a one dimensional parameter that measures the position on the "Z": S$_{Z}$ \cite[][and
references therein]{dieters00}. 
Strong evidence shows that the mass accretion rate $\dot{\textit{M}}$ of individual Z sources 
increases from the top-left to the bottom-right of the "Z" pattern \citep{hasinger90}, i.e. along the 
HB, NB and FB.
Hence, the S$_{Z}$ parameter is supposed to be monotonically related to the mass accretion rate.

For each JEM--X data point (100 s bin and one ScW bin) in the HIDs, the two projections onto both branches 
and the distance to each branch were computed. The branch at the shortest distance 
was chosen and after confirmation via visual inspection, to ensure that data points did not jump
from branch to branch outside the vertex area, the value of S$_{Z}$ was
assigned. 
The scale was set so that the HB-NB vertex was at S$_{Z}=1$ while S$_{Z}=0$ and S$_{Z}=2$ were assigned at 
a 3--12\,keV intensity of 40 counts/s (beginning of HB and end of NB respectively). 
In this frame, S$_{Z}$$<$1 means the source is on the HB (lower $\dot{\textit{M}}$) and 
 1$<$S$_{Z}$$<$2 means that the source is on the NB (higher $\dot{\textit{M}}$).

\subsection{Energy Spectral Modeling}

We averaged all the ISGRI and JEM--X data to 
achieve the best statistics and tried to find the most physically 
reasonable spectral model to fit the entire data set. 
The ISGRI (average) spectrum was obtained from the mosaic image 
shown in Fig.~\ref{fig:ima}, right panel, and the JEM--X total spectrum by averaging the 
44 JEM--X ScW spectra. In this case, the JEM-X spectrum was further grouped to have a 
number of energy bins comparable to ISGRI. 

The X-ray spectra of bright LMXRBs hosting a NS are generally described as the sum of a soft and a hard
component. Two different models are often adopted to describe this composite emission. They are the so-called
 \textit{eastern} model
\citep{mitsuda84} and \textit{western} model \citep{white86}.

 In the eastern model the softer part of the spectrum is a multi-colour disc  describing the emission from 
 the optically-thick, geometrically-thin accretion disc  \citep[XSPEC DISKBB model,][]{mitsuda84}.
  The temperature at the inner disc radius kT$_{in}$ 
 and the normalisation, linked to the inner disc
 radius itself, are the parameters of the model.
  The harder part of the spectrum is a higher temperature Comptonised blackbody 
describing the emission from the NS boundary layer  \citep[COMPBB model,][]{nishimura86}. 
Photons from the NS surface 
at a temperature kT$_{bb}$ are up-scattered by 
 a plasma of  temperature kT$_{e}$ and optical depth $\tau$. In this case the normalisation 
 is linked to the seed photon emitting area.

 In the western model a single temperature blackbody (from the optically thick boundary layer) 
 describes the soft part of the spectrum (BB model in XSPEC). 
 Comptonisation of soft seed photons (kT$_{0}$) in the innermost
 region of the accretion disc
  by a corona of given  temperature kT$_{e}$ and optical depth $\tau$ describes the hard part \citep[e.g. COMPTT model]{titarchuk94}.

\begin{figure}
\centering
\includegraphics[width=1.0\linewidth]{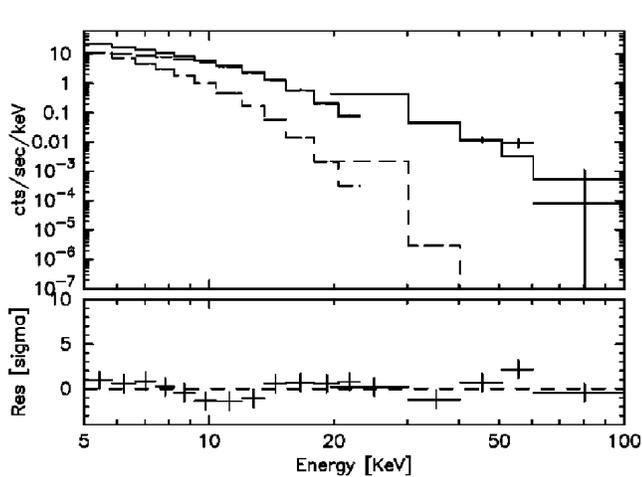}
%\vspace{4cm}
\caption{Best fit of the average ISGRI and JEM--X spectra using the eastern model. The
parameters of the fit are given in Table~\ref{tab:eastwest}. 
Residuals in terms of $\sigma$ and the single spectral components are shown.
 \label{fig:models1}}
\end{figure}

\begin{figure}
\centering
\includegraphics[angle=270,width=1.0\linewidth]{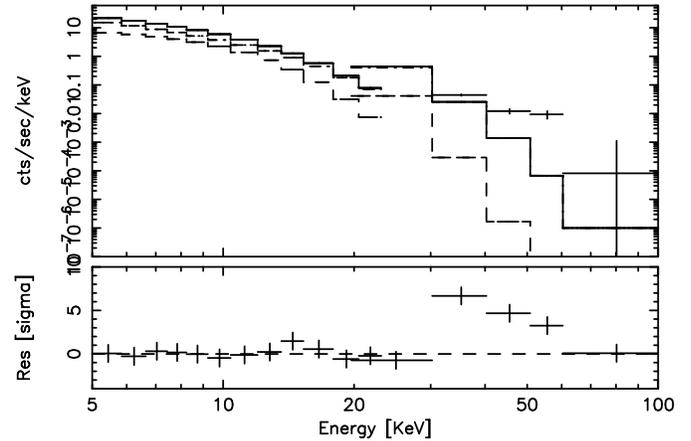}
%\vspace{4cm}
\caption{Best fit of the average ISGRI and JEM--X spectra using the western model.
Residuals in terms of $\sigma$ and the single spectral components are shown. 
A hard excess above $\sim$30\,keV is visible. The
parameters of the fit are given in Table~\ref{tab:eastwest}.
 \label{fig:models2}}
\end{figure}

We used the eastern and western model to fit the average JEM--X and ISGRI spectra simultaneously, and
compared the results. In both cases a galactic absorption by a fixed 
hydrogen equivalent column density of N$_{{\rm H}}$ = 3$\times$10$^{22}$ cm$^{-2}$  \citep{asai94} was added.
\begin{table}
  \begin{center}
    \caption{Best fit parameters for the average ISGRI and JEM--X spectra of GX 5--1. F$_{5-20\,\mathrm{keV}}$ and
    F$_{20-100\,\mathrm{keV}}$
    are the unabsorbed fluxes  
    in the 5-20\,keV and 20-100\,keV range respectively, in units of \rm\,erg\,s$^{-1}$\,cm$^{-2}$; 
    d.o.f. = degrees of freedom. The indicated errors are at 1$\sigma$. No error means the parameter was fixed to the 
    indicated value. The cross-calibration factor was frozen to 0.9 for ISGRI with respect to JEM-X.}
    %\vspace{1em}
    \renewcommand{\arraystretch}{1.2}
    \begin{tabular}[h]{lll}
      \hline
       & Eastern model &  Western model    \\
      \hline       
kT$_{in}$ or kT (keV) & $1.4$ & $1.7$ \\
DISKBB or BB norm & 288  & 0.1\\
kT$_{bb}$ or kT$_{0}$ & $1.93$$\pm0.01$\,keV & $0.87$$\pm0.02$\,keV\\
kT$_{e}$ & $10$\,keV & $3.02$$\pm0.02$\,keV\\
 $\tau$ & $0.37$$\pm0.01$ & $4.15$$\pm0.06$\,keV\\
COMPBB/COMPTT norm &136 & 1.6 \\
Red. $\chi$$^2$ &$1.12$ ($14$ d.o.f.) & $ 6.21 $($13$ d.o.f.) \\
F$_{5-20\,\mathrm{keV}}$ &$1.54$$\times$10$^{-8}$ & $1.54$$\times$10$^{-8}$ \\
F$_{20-100\,\mathrm{keV}}$ &$3.41$$\times$10$^{-10}$ & $3.10$$\times$10$^{-10}$\\ \hline
      \end{tabular}
    \label{tab:eastwest}
  \end{center}
\end{table}
The results of the fit are given in Table~\ref{tab:eastwest} and  Fig.~\ref{fig:models1} and
~\ref{fig:models2}.

\begin{figure}
\centering
\includegraphics[width=1.0\linewidth]{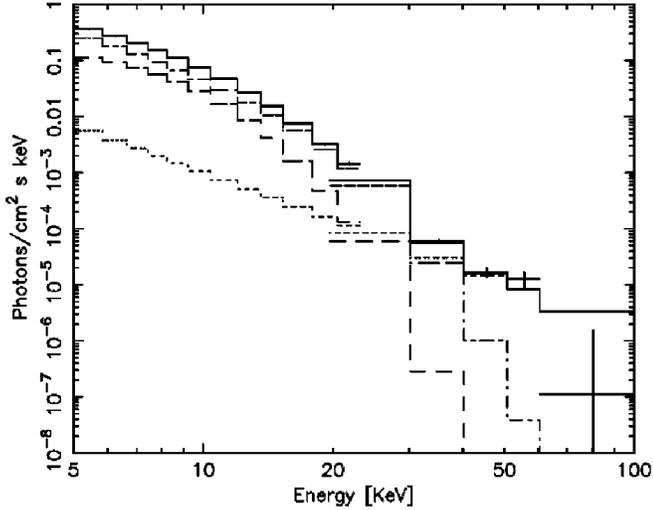}
%\vspace{4cm}
\caption{Photon spectrum of GX 5--1 obtained with the western plus additional power-law ($\Gamma$=2.5) model (Red.
$\chi$$^2$=1.2, 12 d.o.f.).
\emph{Dash}: NS boundary layer emission (BB). \emph{Dot-dash}: Comptonisation of innermost region of the 
accretion disc (COMPTT). \emph{Dot}: additional power-law. \emph{Solid}: total spectrum.
 \label{fig:poufs}}
\end{figure}

\begin{figure}
\centering
\includegraphics[width=0.7\linewidth,angle=270]{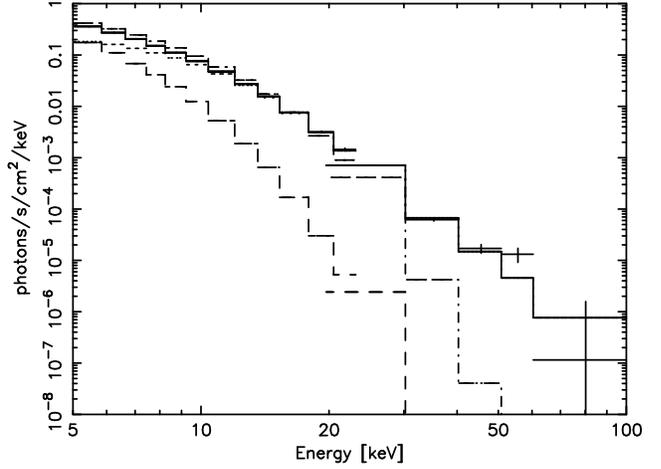}
%\vspace{4cm}
\caption{Photon spectrum of GX 5--1 obtained with the eastern  model described in Table~\ref{tab:eastwest}.
\emph{Dash}: soft, disc blackbody component. \emph{Dot-dash}: hard, NS surface blackbody component \emph{without}
Comptonisation ($\tau$=0). A hard excess is evident.  \emph{Dot} (basically overlapping with the solid line):
NS blackbody component to which  Comptonisation from 
a 10\,keV, $\tau$=0.37 plasma was added.
 \emph{Solid}: total spectrum (disc component 
and NS Comptonised component).
The latter two are basically overlapping above 10\,keV i.e. almost all the emission is coming
from Comptonisation of the NS surface photons.
 \label{fig:ufs}}
\end{figure}

Both models can describe the spectral shape below $\sim$30\,keV equally well.
However, in the western model, there is a significant residual in the ISGRI data above $\sim$30\,keV, that 
 is reasonably well described in the eastern model.  This is because the western model predicts an
exponential cut-off above $\sim$10\,keV, while the eastern model describes the spectral
``flattening'' above $\sim$20\,keV as a Comptonised hard-tail emission. 
We are not able to constrain the Comptonising plasma temperature (and consequently the cut-off energy) from our average
spectrum. In the eastern model, a
good fit can be obtained with a plasma temperature up to about 50\,keV. Either the cut-off  is above the fitted energy range 
or our data are not of high enough quality to constrain the spectral break.
 We choose to fix the temperature to 10\,keV as 
in this frame we can explain in a coherent way the spectral evolution of GX 5--1 (see below).
 
Adding a power-law component with fixed photon index  $\Gamma$$=$$2.5$ in the western model 
results in a good fit 
of the observed hard-excess as shown in Fig.~\ref{fig:poufs}. In the 5--100\,keV band the 
 power-law component alone contributes to about 1.5\% of the total luminosity.
 
Figure~\ref{fig:ufs} shows the photon spectrum obtained with the eastern model parameters shown in 
 Table~\ref{tab:eastwest}.
Since the eastern model provides a physical interpretation for the hard tail, we  focus 
 on this model for the analysis and discussion of the spectral changes of GX 5--1.

\subsection{Spectral variability along the "Z"}

\begin{figure*}
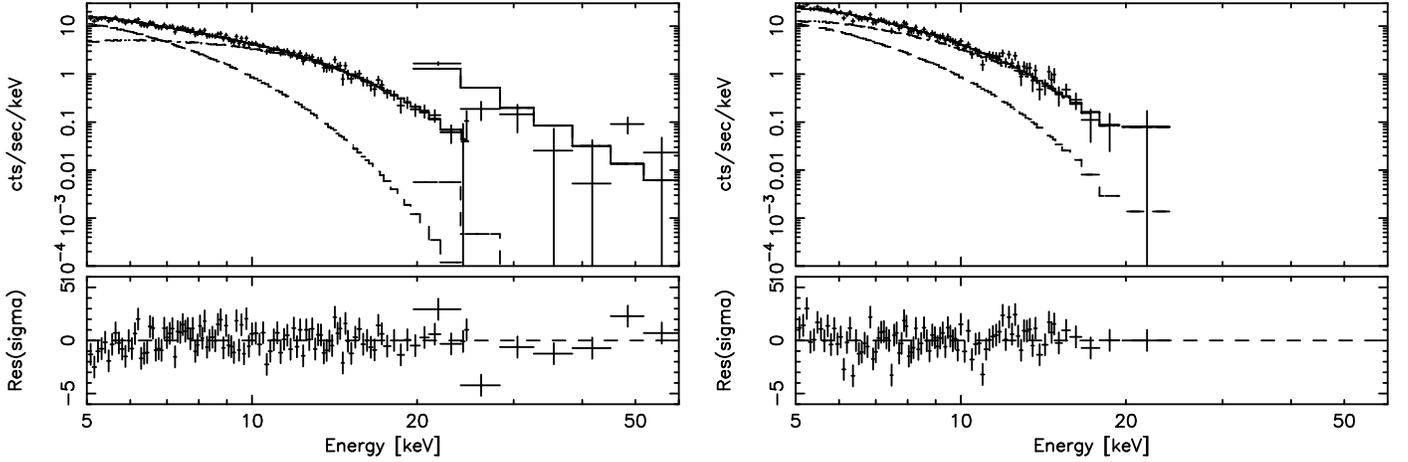

\centerline{
\begin{tabular}{cc}
\psfig{file=./figure9a.ps,angle=270, width=0.5\linewidth} &
\psfig{file=./figure9b.ps,angle=270,  width=0.5\linewidth} 
\end{tabular}}
\caption{ISGRI and JEM--X spectra for GX 5--1 in two different ScWs. The best fit and 
residuals in terms of $\sigma$  are shown. The two components of the model (DISKBB and COMPBB) are also shown (dashed lines).
The Comptonising component is frozen to kT$_{e}$=10\,keV and $\tau$=0.4.
\emph{Left panel}: ScW 012000950010 (IJD=1377.57, UTC=2003-10-09). For this ScW S$_{Z}$=0.50,
i.e. GX 5--1 is in the HB. kT$_{bb}$=2.35\,keV,  red. $\chi$$^2$=1.4 
(112 d.o.f.).  (The residuals obtained in the two ISGRI 
bins between 20--30\,keV are not significant as they are due to known ISGRI systematics). \emph{Right panel}: 
ScW 012200110010 (IJD=1381.63, UTC=2003-10-13). 
S$_{Z}$=1.56 i.e. GX 5--1 is in the NB. The spectrum is softer, only one ISGRI ($>20$\,keV) data 
point met the requirement of
a minimum of 20 counts per bin. kT$_{bb}$=1.49\,keV, red. $\chi$$^2$=1.5 
(89 d.o.f.)  \label{fig:compttscws}}
\end{figure*}

\begin{figure*}
\centerline{
\begin{tabular}{cc}
\psfig{file=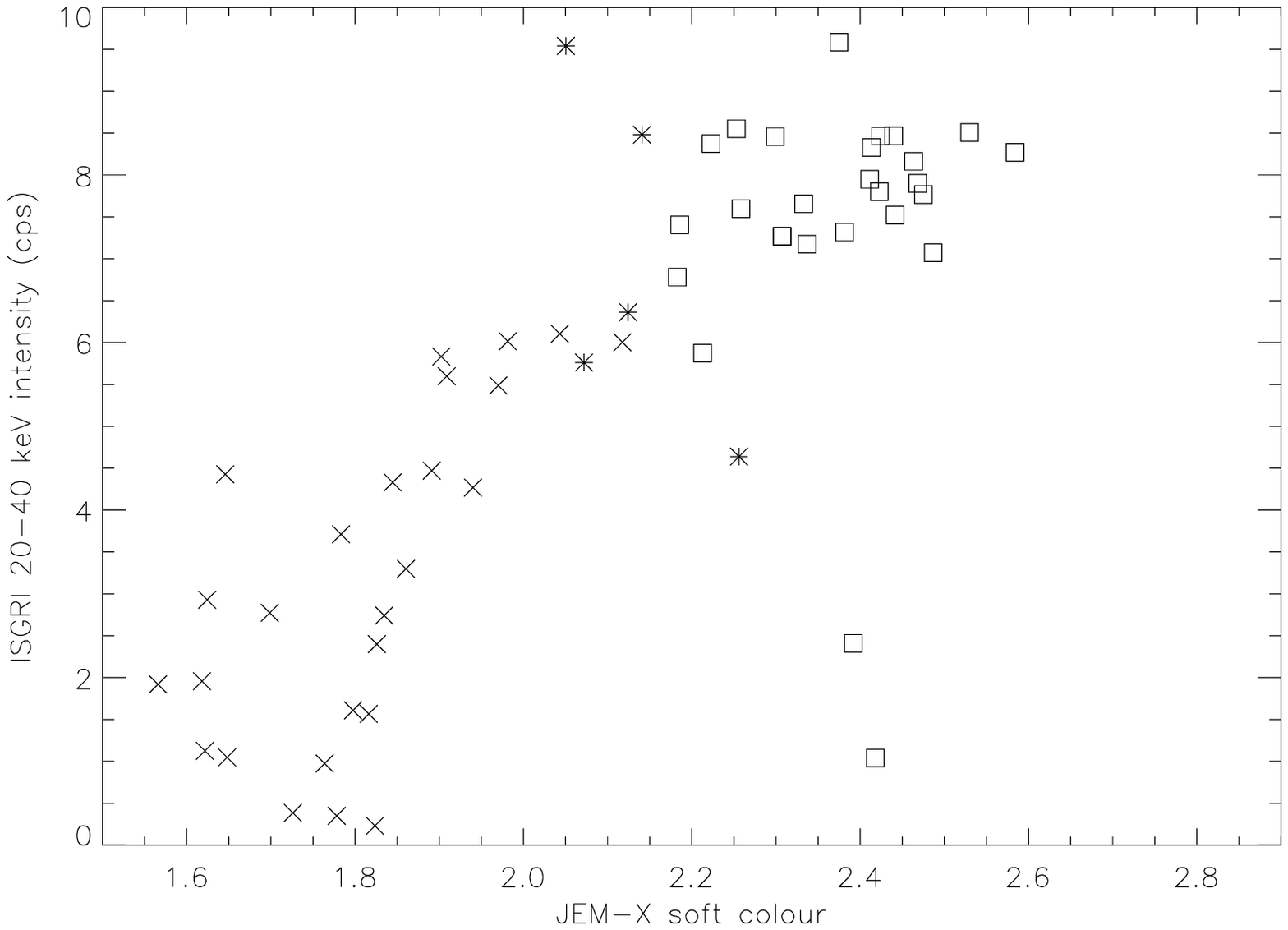, width=0.5\linewidth} &
\psfig{file=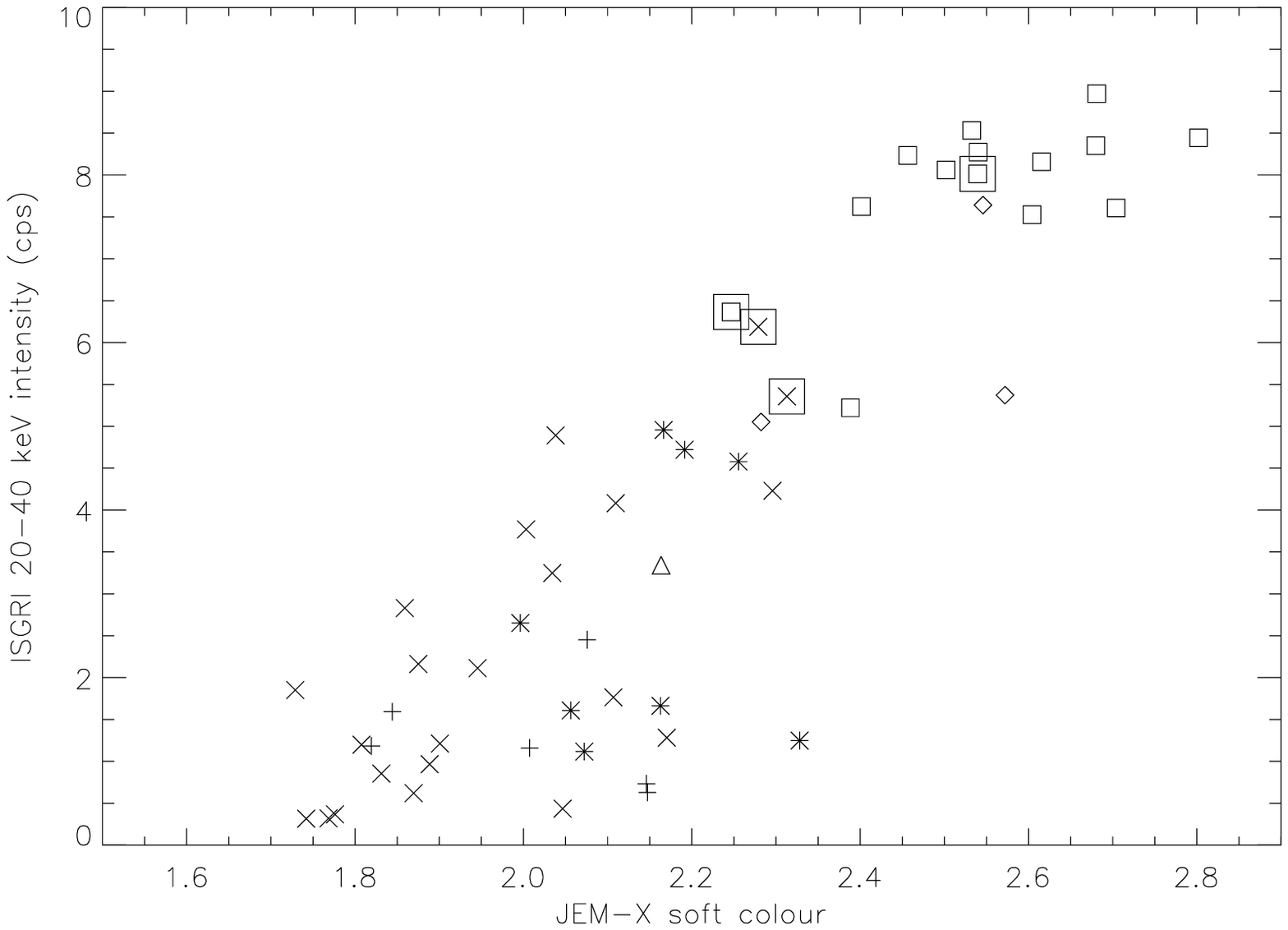, width=0.5\linewidth} 
\end{tabular}}
\caption{Correlation among ISGRI 20--40\,keV flux and JEM--X soft colour, intensity ratio (5--12\,keV)/(3--5\,keV),
for the two datasets (April 2003 on the left and October 2003 on the right).  Each point is one ScW.
 The different symbols used refer to the position of the source in the "Z" track: $\Box$ = clearly HB, 
 $\Diamond$ = uncertain HB, $\triangle$ = HB or NB, $\times$ = clearly NB, $\ast$ = uncertain NB, $+$ = NB or FB. 
 The big boxes indicate transitions between HB and NB (vertex S = 1). The two outliers in the left panel 
 (assigned HB with low ISGRI intensity) are possibly due to systematics.
 \label{fig:jemxisgri}}
\end{figure*}

We studied the spectral variation of GX 5--1 along the HID using the $S_Z$ parameter within the frame of the eastern model (with one 
ISGRI - JEM--X combined spectrum per ScW for a total of 44 ScWs). The soft component (DISKBB) was frozen to the values obtained
in the best fit given in Table 1, for all the ScWs.
Similarly, the plasma temperature kT${_e}$ was fixed at a value of 10\,keV, since this parameter is strongly correlated
with $\tau$. The normalisation of the hard component, kT$_{bb}$  and $\tau$ were let free to vary from one Scw fit to the next
as well as the instrument cross-calibration factor.
An average value of reduced $\chi$$^{2}$$\simeq$1.2 was obtained. We found a clear relation 
between the S$_{Z}$ value and the deduced kT$_{bb}$ for all the ScWs. 
To see if we could describe the spectral evolution of GX 5--1 in a more constraining way, we proceeded fitting the data
letting in one case  kT$_{bb}$ vary (with $\tau$ fixed to 0.4) and in the other 
 $\tau$ vary (kT$_{bb}$ fixed to 2.0\,keV). In both cases the normalisation was let free as well.
Fixing  kT$_{bb}$ resulted in a much worse fit (average reduced $\chi$$^{2}$$\simeq$2.5) while 
fixing  $\tau$ gave an equally acceptable fit (average reduced $\chi$$^{2}$$\simeq$1.3). This is consistent with the fact that 
 kT$_{bb}$ has a clear relation with the position of GX 5--1 along the "Z" and should not be frozen. Conversely, $\tau$ shows a random 
 distribution of values around 0.4 (extreme cases being 0.1 and 0.8), it is not the main parameter driving the spectral evolution along
 the "Z" and hence can be fixed.\\
The combined JEM--X and ISGRI spectral fitting for two ScWs, one for the
HB and the other one for the NB, are shown in Fig.~\ref{fig:compttscws}, left and right respectively. 
The spectrum extracted from the HB is significantly harder than the NB one. This is true also for the 
other spectra and the hard X-ray emission is indeed characteristic of the HB. In fact we obtain 
that the hard X-ray emission (ISGRI countrate) is systematically linked to the position of the source in the CC diagram: 
Fig.~\ref{fig:jemxisgri} shows the correlation among ISGRI 20--40\,keV flux and JEM--X soft colour for the two
datasets (off-axis angle $<$$5^\circ$). 
The different symbols used refer to the position of the source in the "Z" track: the cases
that have been identified with the HB  have a higher ISGRI flux whereas the NB cases 
have a lower ISGRI flux.
The second dataset, Fig.~\ref{fig:jemxisgri} right panel, has more uncertain cases as expected 
given that its "Z"
shape is less clean than for the first dataset.

\begin{figure*}
\centerline{
\begin{tabular}{cc}
\psfig{file=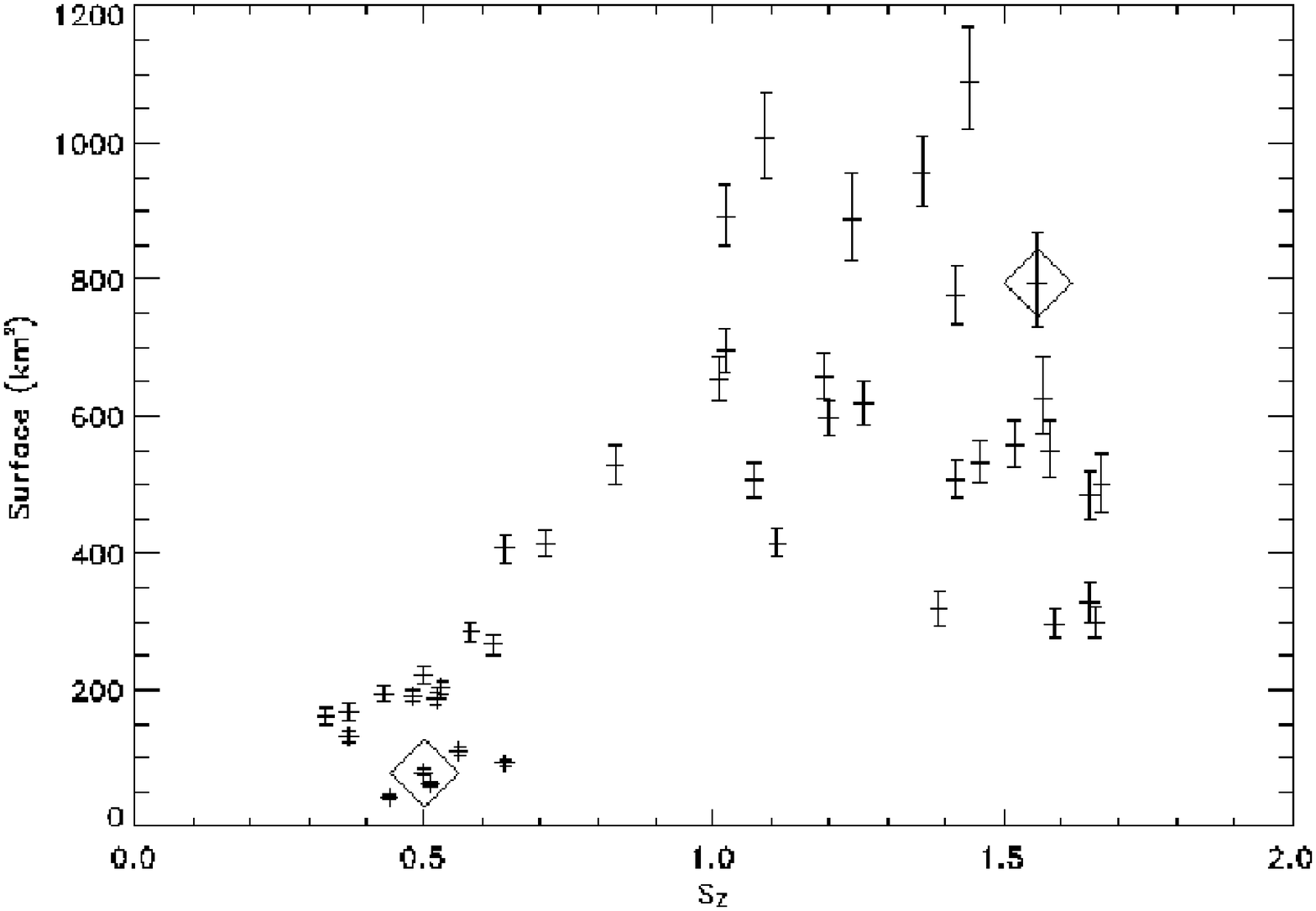, width=0.5\linewidth} &
\psfig{file=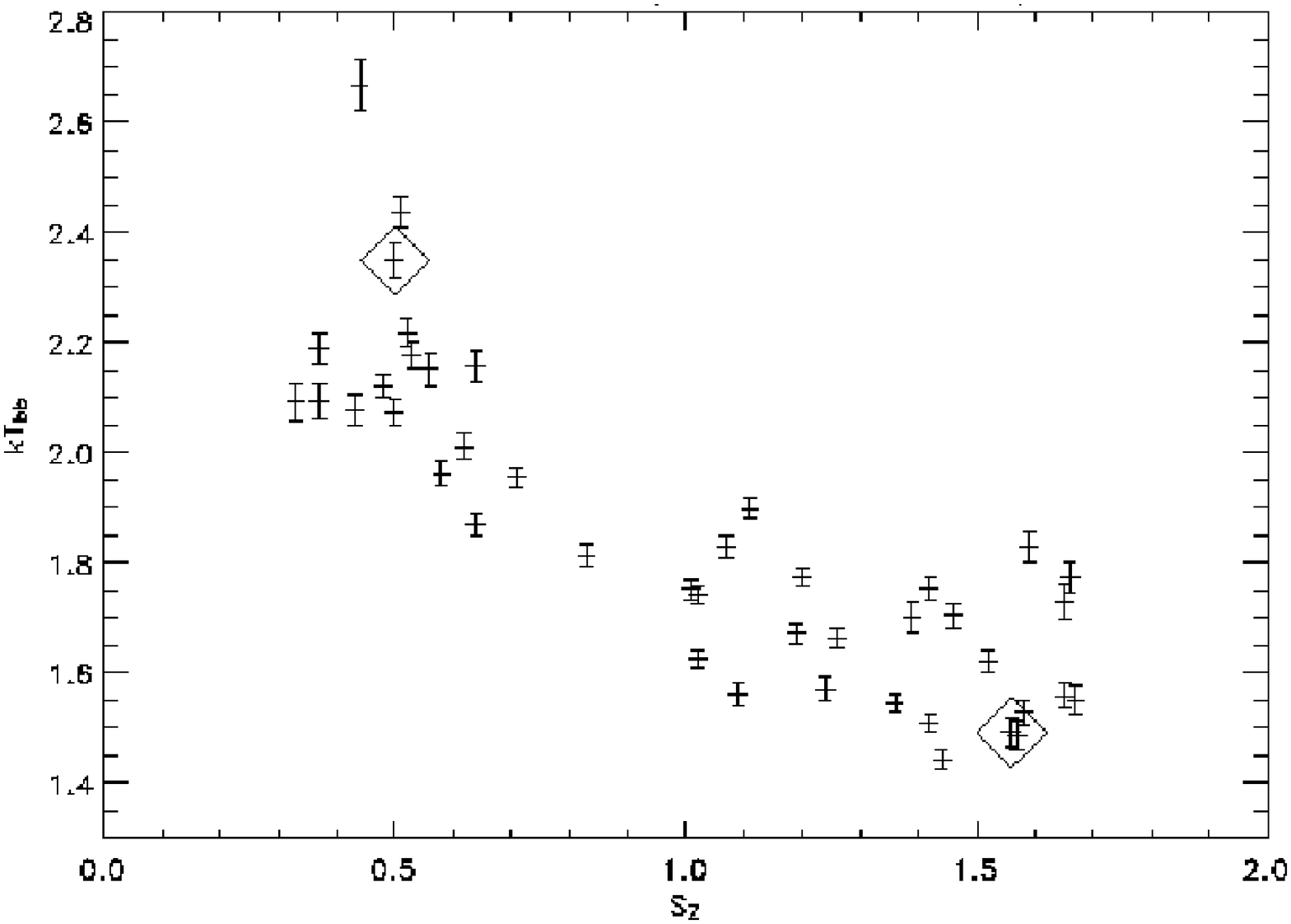,  width=0.5\linewidth} 
\end{tabular}}
\caption{Relations among the seed photons' properties and position of the source in the "Z". The 
electron plasma temperature kT${_e}$ and the optical depth were frozen to 10\,keV and 0.4 respectively. 
In both panels the two ScWs of Fig.~\ref{fig:compttscws} are highlighted. \emph{Left panel}: 
  relation between 
the  blackbody emitting surface 
and  S$_{Z}$. The surface is in units of km$^{2}$ and is computed as $\pi$$\times$$d$$^{2}$$\times$normalisation, 
where $d$ is the distance of the source in units of 10\,kpc, assumed to be 0.8. 
%This relation has a correlation coefficient of R=$0.54$ in Spearman statistics with a probability of having the 
%same correlation with a randomly distributed sample of P$_{rand}$$\simeq$10$^{-4}$.
\emph{Right panel}:
relation between the derived  blackbody 
temperature kT$_{bb}$ and S$_{Z}$. 
% In this case the correlation coefficient was of R=$-$0.81  with
%P$_{rand}$$\simeq$10$^{-11}$.
\label{fig:final}}
\end{figure*}

 Figure~\ref{fig:final}
shows the variation of the emitting hot NS surface area and temperature, kT$_{bb}$,
along the "Z" (left and right panel respectively).  
While GX 5--1 moves from the HB to the NB (left to right in 
Fig.~\ref{fig:final}), the size of the emitting surface increases while its temperature decreases. 

\section{Discussion}
Our observations constitute the first opportunity in which the Z source GX 5--1 could 
be monitored in a long term period in the hard X-ray domain without contamination
from the nearby BHC \mbox{GRS 1758-258}. \\
The source was thoroughly covered by \textit{INTEGRAL} mainly
during  two  periods (April and October 2003).
The HIDs of \mbox{GX 5--1} for the two periods show a clear secular shift. This is consistent with
previous observations of GX 5--1 itself as well as of other Z sources 
\citep[][and references therein]{kuulkers94, kuulkers95}  for which a secular shift was detected.
In our data, GX 5--1 covers widely the HB 
and NB while very poor coverage of the FB (if any) is seen, similarly to previous observations
\citep[e.g.,][]{kuulkers94}.

\subsection{Hard X-ray Emission}

The  ISGRI and JEM--X average spectra show  that 
\mbox{GX 5--1} has a clear hard X-ray emission above 20\,keV. 
Assuming a distance of GX 5--1 of 8\,kpc, we obtain luminosities of the order of 
L$_{1-20\,\mathrm{keV}}$$\sim$10$^{38}$\rm\,erg\,s$^{-1}$ (\footnote{Obtained extrapolating our 
\textit{INTEGRAL} best fit model down to 1\,keV.}) and L$_{20-200\,\mathrm{keV}}$$\sim$10$^{36}$\rm\,erg\,s$^{-1}$,
fully consistent with values obtained for other Z sources \citep{disalvo02}. 
With such luminosities, similarly to the case of the Z sources \mbox{GX 17+2} and \mbox{GX 349+2}, 
\mbox{GX 5-1} lies outside the so-called burster box \citep{barret00},
in a region that was originally believed to be populated by BH LMXRBs. Like BHs, NS LMXRBs can have a hard X-ray emission 
when they are bright. 

At least in the cases in which a cut-off is observed, 
the hard X-ray emission can be interpreted as thermal Comptonisation of soft photons in a hot, 
optically thin plasma in the vicinity of the neutron star. In the case of GX 5--1
 we are not able to constrain the Comptonising plasma temperature. This suggests that either 
 our data are not of good enough  quality to measure the cut-off or that the cut-off is beyond the fitted energy range. 
 In any case, we do obtain a good fit with relatively low plasma temperatures (10\,keV) hence  non-thermal
 processes, even if not ruled out, cannot be claimed.
 
In cases where the cut-off has not been observed up to about 100\,keV or 
higher \citep{disalvo00,disalvo01}, extremely high electron temperatures are required, which is 
unlikely, given that the bulk of the emission of Z sources is very soft. 
Non attenuated power-law tails dominating the spectra at high energies may
be produced by  mildly relativistic ($v/c \sim 0.1$) outflows 
with flatter power laws corresponding to higher optical
depth of the scattering medium and/or higher bulk electrons velocities,
in a way that is similar to thermal Comptonisation \citep{psaltis01}. 
Alternatively, X-ray components could originate from 
the Comptonisation of seed photons by the non thermal high energy-electrons of a  jet \cite[][and 
references therein]{disalvo02} or via synchrotron self-Compton emission
directly by the jet \citep{markoff04}.

\subsection{Spectral Variation}

With \textit{INTEGRAL}, we are able to
study the spectral state of \mbox{GX 5--1} within one ScW seeing the source variability above 20\,keV that has never been 
studied before. On the other hand, the softer part of the spectrum, dominated by 
the disc emission, cannot be constrained and thus we fixed it
to the value found from the JEM-X average spectrum.

The spectral changes of GX 5--1 along the "Z" can be explained by the smooth  
variation of the physical properties of the seed photons originating from the hot NS surface (heated by the boundary layer).
This scenario is different from the more complicated  ones that e.g. \cite{hasinger90} and \cite{hoshi91}   proposed to
explain \textit{GINGA} data of Cyg X--2 and GX 5--1 respectively. 
\cite{hasinger90} studied the spectral variations of Cyg X-2 using both the eastern and western model.
The temperature and the luminosity of both components (soft and hard) were let free to vary in both models and a
rather complicate interplay of the parameters of both components was describing the spectral variations of the source along 
the "Z" pattern. 
\cite{hoshi91} used the eastern model for GX 5--1 and the spectral changes were also in this case 
due to the variation of different parameters: in the NB the inner disc 
radius temperature  kT$_{in}$ and the NS boundary layer temperature kT$_{bb}$ remained constant and spectral shape changes 
were explained by an increase of the emission area (from lower left to upper right NB). 
While the spectrum became harder, in the HB, an additional parameter, kT$_{in}$, was let free to reproduce the trend 
of the data. 

Unlike \textit{GINGA}, with \textit{INTEGRAL} data we cannot constrain the variability of 
the disc component and thus we had to fix it to the values found from the average spectra.
 The normalisation of this component (modelled with DISKBB in XSPEC terminology) led to an inner disc radius of 
R$_{in}$(cos$\theta$)$^{1/2}$ $\simeq$14(cos$\theta$)$^{1/2}$\,km for an assumed source distance of 8\,kpc where
$\theta$ is the angle between the normal to the disc and the line of sight. In the inner region of the disc, 
a hot outer
layer  affects the emergent spectra by Comptonisation. This leads to an observed temperature T$_{col}$ which
is higher than the effective temperature T$_{eff}$. As a consequence the observed inner disc radius 
is underestimated by a factor of
(T$_{col}$/T$_{eff}$)$^{2}$. With this hardening factor correction assumed to be 1.5 \citep{ebisawa91}, we obtained 
an inner disc of about R$_{eff}$(cos$\theta$)$^{1/2}$$\simeq$30(cos$\theta$)$^{1/2}$\,km.

 Not being sensitive to changes in the softer part of the spectrum is a limitation in our 
study as we do expect the soft part of the spectrum (the disc) to vary but, for the first time, 
we were able to study the variability of GX 5--1 in the less explored hard X-rays, above 20\,keV. We
see that the hard component is indeed chatacteristic of the HB, namely the spectrum is
harder at lower accretion rates. This seems to be the general trend in LMXRBs \citep{barret94}.  
Atoll sources are thought to be at a lower level of accreting rate than Z sources and 
indeed they are generally harder. This trend continues also within a class of sources:
the hard components observed in the Z sources GX 17+2 \citep{disalvo00}, Cyg X-2 \citep{disalvo02b} and GX349+2 \citep{disalvo01}
decrease in intensity from HB to NB and FB, i.e. for increasing mass accretion rate. An exception to this trend 
was found in Sco X--1 \citep{damico01} where the presence of the hard component was not
related to the position of the source in the "Z" track.
Theoretical interpretations for this general 
trend have already been discussed \citep{inogamov99, popham01, kluzniak85}.
 
\cite{popham01} have computed the boundary layer structure and its evolution
 with mass accretion rate $\dot{\textit{M}}$. They found a significant dependence 
of the boundary layer size (both height and radial extension) with  $\dot{\textit{M}}$. 
Matter falls on a portion of the NS surface that is touched by the boundary layer. 
This part of the NS surface is thermalised and emits blackbody emission.  As $\dot{\textit{M}}$ increases, 
the area covered by the boundary layer will increase and so will the NS surface responsible of the 
blackbody emission.
This is the same trend we found in the case of GX 5--1 (Fig.~\ref{fig:final}, left): S$_{Z}$, i.e. $\dot{\textit{M}}$,
increases and the emitting NS surface increases. This trend is more clearly visible in the HB than in the NB.
 A possible explanation for this could be that along the NB (with increasing mass accretion rate), 
the accretion disc, i.e. the soft part of the spectrum that we are not able to constrain with \textit{INTEGRAL}, is starting to change.
 Besides, in the NB $\dot{\textit{M}}$ is not expected to vary much (as it is already close to $\dot{\textit{M}}_{Edd}$) so
this may be also (part of) the reason why there is no clear trend in the NB. 

The 5--100\,keV observed luminosity
reached a maximum variation of a factor of $\sim$3, unabsorbed 
L$_{5-100\,\mathrm{keV}}$$\simeq$(0.6-1.7)$\times$10$^{38}$\rm\,erg\,s$^{-1}$ (\footnote{This range includes the 10\% uncertainty 
due to the current calibration status.}). 
This is small compared to the NS emitting surface variation (factor of $\sim$20, see Fig.~\ref{fig:final} left). L$_{5-100\,\mathrm{keV}}$ is the total luminosity
obtained from the composition of soft and hard component between 5--100\,keV but 
the variation in L$_{5-100\,\mathrm{keV}}$ is to be attributed to the changes in 
the hot NS surface properties alone as  the boundary layer dominates the emission and the disc as well as the parameters of
the Comptonising plasma were kept fixed. In our interpretation, any variation in L$_{5-100\,\mathrm{keV}}$ is a consequence of 
changes in the properties of the hot NS surface that we modelled with a blackbody emission. Its luminosity is proportional
to the emitting surface and temperature, L$_{bb}$ $\propto$R$^{2}$$\times$T$_{bb}$$^{4}$.
So, if the luminosity stays nearly constant (factor of $\sim$3) while the emitting surface
increases (factor of $\sim$20), then the temperature of the emitting surface will decrease.
 Thus, the anti-correlation that we found between  S$_{Z}$
and the blackbody temperature kT$_{bb}$ (Fig.~\ref{fig:final}, right) may be qualitatively understood.

\section{Conclusions}
We have studied the X-ray emission of the Z source \mbox{GX 5--1} with \textit{INTEGRAL}. 
A clear hard emission above $\sim$20\,keV is detected and it may have the same origin of the 
hard components observed in other Z sources as it shares the property of being characteristic to the HB. 
We used the so-called ``eastern'' and ``western'' model to fit the average spectra of \mbox{GX 5--1}. The eastern
model  describes the spectral
``flattening'' above $\sim$20\,keV as a Comptonised hard-tail emission whereas the western one predicts an
exponential cut-off above $\sim$10\,keV and an additional power-law is needed to account for the hard X-ray emission.
Since the eastern model provides a physical interpretation for the hard tail, we  focussed 
 on this model for the analysis of the spectral changes of GX 5--1.

We interpret the spectral changes of GX 5--1  along the "Z" pattern  in the 
hardness intensity diagram in terms of Comptonisation of varying soft photons ($\sim$2\,keV)
by a hot plasma (10\,keV). The soft photons are interpreted as blackbody emission from 
the part of the NS surface that is heated by the boundary layer.
The Comptonising plasma is the boundary layer optically thin plasma. 
When GX 5--1 moves downwards in the "Z", 
the temperature and optical depth of the Comptonising   plasma 
 are not seen to  change in the data. What changes along the "Z" is the temperature of the optically thick 
 emission from the hot NS surface that shows a steady decrease with increasing 
 mass accretion rate. 
This  may be a consequence of the
gradual expansion of the boundary layer that we detect in the data.
This trend  is in agreement with
 theoretical studies \citep{popham01} that indeed predict the expansion
of the boundary layer surface with increasing mass accretion rate.

With the  \textit{INTEGRAL} long term monitoring we will also have higher chances to catch GX 5--1 in the FB.
 The  behaviour of Z sources 
on the FB, believed to correspond to super-Eddington accretion, is quite complex and it will be interesting to see 
 if it can be smoothly included in the HB and NB scenario described here.
Furthermore we will search 
for  hard tails extending above 100\,keV, in GX 5--1 as well as in the other Z sources and NS binaries of our
monitoring program.

The synergy of a long term monitoring and a multi-wavelength 
study should be the best way to  understand the physics of hard-X ray emission
of LMXRBs.  Our collaboration is moving in this direction: regular 
coordinated \textit{INTEGRAL}-\textit{RXTE} observations
of a sample of Atoll and Z sources (including GX 5--1) are being 
performed and simultaneous radio and optical observations are  attempted.

\begin{acknowledgements}
We aknowledge the anonymous referee for useful comments. A.P. acknowledges V. Beckmann and L. Foschini for careful reading of the manuscript and useful discussion.
J.R. acknowledges M. Cadolle-Bel for great help with
ISGRI spectral responses.
\end{acknowledgements}

\bibliographystyle{aa}
\bibliography{biblio}

\end{document}